\documentclass{ws-procs975x65}

\begin{document}

\title{The healthy DGP branch in a state of self-acceleration}

\author{Mariam  Bouhmadi-L\'opez$^1$} 
\address{Centro Multidisciplinar de Astrof\'{\i}sica - CENTRA,
 Departamento de F\'{\i}sica,\\ Instituto Superior T\'ecnico,
Av. Rovisco Pais 1, 1096 Lisboa, Portugal\\ E-mail: \mbox{$^1$mariam.bouhmadi@ist.utl.pt}}

\begin{abstract}
We present a method to self-accelerate the normal DGP branch. We obtain the maximally symmetric solutions and study their stability.
\end{abstract}


\bodymatter

\section{Motivation}

Despite the huge efforts to understand the current acceleration of the universe, no satisfactory answer has been provided so far for this puzzling speed up of the universe. The mainstream are based in invoking a dark energy component, a cosmological constant or a modified theory of gravity (at the appropriate scales) \cite{Durrer:2008in,Nojiri:2006ri,Capozziello:2007ec}. 

In this short note, we will focus on the latter approach. More precisely, we will suggest a way to self-accelerate the normal Dvali, Gabadadze and Porrati  (DGP) branch \cite{Dvali:2000hr} which is known to be free from the ghost problem present on its twin self-accelerating DGP branch. The approach is quite simple and is based in modifying the brane action by including further corrections to the Hilbert-Einstein action of the brane. For simplicity, we will choose those corrections/modifications to be modelled by a function of the scalar curvature of the brane\cite{Nojiri:2006ri,Capozziello:2007ec}. 

In the next section, we will present our model. Then we will obtain the self-accelerating solutions corresponding de Sitter branes and we will analyse their stability under homogeneous perturbations. We will conclude with some remarks.

\section{The setup}

We consider a  brane, described by a 4D hyper-surface ($h$, metric g), embedded in a 5D bulk space-time ($\mathcal{B}$, metric  $g^{(5)}$), whose action is given by 
\begin{eqnarray}
\mathcal{S} = \,\,\, \int_{\mathcal{B}} d^5X\, \sqrt{-g^{(5)}}\;
\left\{\frac{1}{2\kappa_5^2}R[g^{(5)}]\;\right\}
 + \int_{h} d^4X\, \sqrt{-g}\; \left\{\frac{1}{\kappa_5^2} K\;+\alpha f(R) + \mathcal{L}_m \right\}\,, \label{action}
\end{eqnarray}
where $\kappa_5^2$ is the 5D gravitational constant,
$R[g^{(5)}]$ is the scalar curvature in the bulk and $K$ the extrinsic curvature of the brane in the higher dimensional
bulk. On the second term of the action,  $R$ is the scalar curvature of the induced metric on the brane, $g$, and  $\alpha$ is a constant that measures the strength of the generalised induced gravity term $f(R)$ and has mass square units. 
Finally, $\mathcal{L}_m$ corresponds to the Lagrangian of matter confined on the brane. We assume a $\mathcal{Z}_2$ symmetry across the brane. The previous action includes as a particular case the DGP model \cite{Dvali:2000hr} for $f(R)=R$ and $\alpha=1/2\kappa_4^2$ where $\kappa_4^2$ is proportional to the 4D gravitational constant. It can be shown that the total energy density of the brane is conserved (we refer the reader to \cite{BouhmadiLopez:2009db} for more details). In particular, the energy density of matter on the brane is conserved.

We will consider a FLRW brane filled with matter with energy density $\rho^{(m)}$ and pressure $p^{(m)}$. The effects of the  $f(R)$ term on the brane action can be described through an effective energy density, $\rho^{(f)}$, and pressure,  $p^{(f)}$, that read
\begin{eqnarray}
\rho^{(f)} &=&-2\alpha\left[3\left(H^2+\frac{k}{a^2}\right)f'-\frac12(Rf'-f)+3H\dot R f'' \right], \label{rhof}\\
p^{(f)}&=& 2\alpha\left\{\left(2\dot H+3H^2+\frac{k}{a^2}\right)f'-\frac12(Rf'-f) \left[\ddot R f''+(\dot R)^2f'''+2H\dot R f''\right]
\right\},\label{pf}
\end{eqnarray}
Notice that the definitions of $\rho^{(f)}$ and $p^{(f)}$ are different from the standard 4D definition in $f(R)$ models \cite{BouhmadiLopez:2009db}. It can be shown that $\rho^{(m)}$ and $\rho^{(f)}$ are both conserved.  Finally, we can write the modified Friedmann equation on the brane (for the flat chart) as 
\begin{eqnarray}
3H^2= \frac{\kappa_5^4}{12}\rho^2.\label{friedmann}
\end{eqnarray}
where the  total energy density $\rho$ and the total pressure $p$ of the brane are defined as  
\begin{eqnarray}
\rho=\rho^{(m)}+\rho^{(f)}, \quad
p=p^{(m)}+p^{(f)}.
\label{rhop}
\end{eqnarray}

\section{The self-accelerating branes and their stability}

We obtain on this section the de Sitter solutions of the model presented on the previous section. Those solutions will correspond to self-accelerating branes; i.e. the de Sitter character of the brane is not due to the presence of matter on the brane. Therefore, in what follows we will assume $\rho^{(m)}=0$ and  $p^{(m)}=0$. 

The Hubble parameter for the de Sitter solutions can expressed as \cite{BouhmadiLopez:2009db} 
\begin{eqnarray}\label{desitterH1}
{2\kappa_5^4\alpha^2F_0^2}H_0^2=1 + \frac13\kappa_5^4\alpha^2F_0(R_0F_0-f_0)+\epsilon\sqrt{1+\frac23\kappa_5^4\alpha F_0\big[\alpha(R_0F_0-f_0)\big]}
\end{eqnarray}
where $\epsilon=\pm 1$, the subscript 0 stands for quantities evaluated at the de Sitter space-time, $R_0=12 H_0^2$ and $F=df/dR$. We recover the DGP model for $f(R)=R$. In fact, in that case, the de Sitter self-accelerating DGP branch is obtained for  $\epsilon= 1$ and the normal DGP branch or the  non-self-accelerating solution for $\epsilon=-1$. When the brane action contains curvature corrections to the Hilbert-Einstein action given by the brane scalar curvature, the branch with $\epsilon=-1$ is no longer flat and accelerates. Therefore, an $f(R)$ term on the brane action induces in a natural way  self-acceleration on the normal branch. Most importantly, it  is known that such a branch is free from the ghost problem. 

We next analyse the stability of de Sitter solutions under homogeneous perturbations up to first order on $
\delta H= H(t)-H_0$. We will follow the method used in \cite{Faraoni:2005vk}. The perturbed Friedmann equation (\ref{friedmann}) implies an evolution equation for $\delta H$:
\begin{equation}
\delta \ddot H + 3H_0\delta \dot H + m_{\rm{eff}}^2\delta H=0,
\label{Friedmannpert}
\end{equation}
where $m_{\rm{eff}}^2$ is defined as
\begin{eqnarray}
m_{\rm{eff}}^2&=&  \frac13\left(\frac{F_0}{f_{RR}}-2\frac{f_0}{F_0}\right) \\
&-&\frac{2}{\alpha^2\kappa_5^4F_0^2}\left[1-\sqrt{1+\frac23\alpha^2\kappa_5^4F_0(f_0-\kappa_5^2UF_0)}\right]\nonumber\\
&+&\frac{F_0}{3f_{RR}}\left[1-\sqrt{1+\frac23\alpha^2\kappa_5^4F_0(f_0-\kappa_5^2UF_0)}\right]^{-1}\nonumber
\end{eqnarray}
where $f_{RR}={d^2f}/{dR^2}$. All these quantities are  evaluated at the de Sitter background solution. Any de Sitter solution is stable as long as $m_{\rm{eff}}^2$ is positive. 

If we assume that we are close to the 4D regime; i.e. the Hubble rate of the brane is close to its analogous quantity in a 4D $f(R)$ model, it can be shown that the extra-dimension has  a \textit{benigner} effect in the 4D f(R) model, as long as  ${F_0^2}<{4f_0} f_{RR}.$ (see \cite{BouhmadiLopez:2009db} for more details).

\section{Conclusions and further comments}

In this note, we have presented a mechanism to self-accelerate the normal DGP branch. Unlike the original self-accelerating DGP branch, it is know that the normal DGP branch is free from the ghost problem.  The mechanism we introduced is based in including  curvature modifications to the Hilbert-Einstein action of the brane. For simplicity, we choose those terms to correspond to an arbitrary function of the scalar curvature of the brane, which in addition is known to be the only higher order gravity theories that avoid the so called Ostrogradski instability in 4D models.

\section*{Acknowledgements}
M.B.L. is supported by FCT (Portugal) through the fellowship FCT/BPD/26542/2006.

\end{document}